\documentstyle[12pt]{article}
\begin{document}

\begin{titlepage}
\begin{flushright}
JINR-E2-94-400  \\  October 1994
\end{flushright}
\vglue 1cm
\begin{center} {\large \bf
RENORMALIZATION GROUP IMPROVED RADIATIVE CORRECTIONS \\
TO THE SUPERSYMMETRIC HIGGS BOSON MASSES}
\vglue 2cm
{\bf A.V.Gladyshev\footnote{Physics Department, Moscow State
University; e-mail: gladysh@thsun1.jinr.dubna.su} and
D.I.Kazakov\footnote{Supported in part by INS Grant \#RFL000; e-mail:
kazakovd@thsun1.jinr.dubna.su}}
\vglue 1cm
{\it Bogoliubov Laboratory of Theoretical Physics, Joint Institute for
Nuclear Research,\\ 141 980 Dubna, Moscow Region, RUSSIA }
\end{center}

\vspace{2cm}

\begin{abstract}
The one-loop radiative corrections to the Higgs boson potential in the
MSSM, originating from the top quark and squark loops, are summed in
the leading log approximation using the renormalization group. The RG
improved effective potential is minimized and the corrections to the
CP-odd and CP-even Higgs boson masses are calculated. The resulting
masses exhibit smoother top mass dependence than those calculated
without RG summation. We have also found that for preferable values of
the top mass the light Higgs mass does not exceed 100 GeV.
\end{abstract}

\end{titlepage}

\section{Introduction}

It has been recently pointed out~\cite{Zwirner,Arnowitt} that the
radiative corrections to the masses of the Higgs bosons in the
framework of the MSSM~\cite{Ross},~\cite{Ibanez},~\cite{Barbieri} can
be relatively large. The leading correction to the effective potential
comes from the top quark and squark loops, being proportional to the
top Yukawa coupling, which is considered to be big due to the heaviness
of the top quark. The other corrections happen to be smaller, though in
some cases their effect is not negligible as well. The net effect of
the radiative corrections is to increase the masses of the Higgs
bosons. This increase may be very significant for the future searches,
since it can achieve several dozen GeV, implying that the Higgs mass
could exceed the $Z$ boson mass.

Considering the tree-level Higgs potential one finds out that the value
of the lightest Higgs boson mass is restricted by the inequality

\begin{equation}
\label{1}m_h<M_Z.
\end{equation}

This strict limit is, however, violated by the radiative corrections.
The radiative corrections to the supersymmetric boson masses proceed
>from the one-loop effective potential

\begin{eqnarray}
V&=&V_{tree}+\Delta V_{1 loop}, \label{2} \\
\Delta V_{1 loop}&=&\frac{1}{64\pi^2}
Str {\cal M}^4 \left(\log\frac{{\cal M}^2}{Q^2}
-\frac{3}{2}\right), \label{3}
\end{eqnarray}
where $Str$ denotes the conventional supertrace and $Q^2$ is the scale
at which all the couplings in the tree-level potential are
renormalized. ${\cal M}$ are the field dependent masses of all the
possible particles running through the loops. In what follows we limit
ourselves with the top and stop contributions as the main ones.

As far as the radiative corrections appear to be large achieving 30 \%
one can wonder about the values of the higher order contributions.
Since according to eq.(\ref{3}) they have the $log$ form one is
expecting to have $log^2$ at the second loop, $log^3$ at the third
loop, etc. Being essential these logs have to be summed giving
considerable change of the results.

Indeed this happens in the simplest case of the $\phi ^4$ model
considered in the pioneering paper by Coleman and
Weinberg~\cite{Coleman}. The summation of the leading logs to the
effective potential changed the situation qualitatively leading to the
disappearance of a non-trivial minimum arising at the one-loop level.

The summation procedure can be naturally done with the help of the
renormalization group technique, which we are going to apply to our
particular case.

\section{RG Improved Effective Potential}

Let us remind the expression for the one-loop effective Higgs potential
in the MSSM which takes into account the radiative corrections due to
the top quark and squark loops. For the neutral Higgses it has been
calculated in ref.\cite{Zwirner} and has the form

$$  V(H_1,H_2) = m^2_1|H_1|^2+m^2_2|H_2|^2-m^2_3(H_1H_2+h.c.)+
\frac{g^2+g^{\prime 2}}{8}(|H_1|^2-|H_2|^2)^2  $$
\begin{equation}
+ \frac{3}{32\pi^2} \left[
\tilde{m}_{t1}^4 ( \ln \frac{\tilde{m}_{t1}^2}{Q^2}-\frac{3}{2})
+\tilde{m}_{t2}^4 ( \ln \frac{\tilde{m}_{t2}^2}{Q^2}-\frac{3}{2})
-2{m}_{t}^4(\ln \frac{{m}_{t}^2}{Q^2}-\frac{3}{2})\right]-V_0
\label{5},
\end{equation}
where $\tilde m_{ti}$ are the field dependent masses of the stop
particles and $m_t$ is the field dependent top mass. The scale $Q^2$
remains arbitrary and is usually chosen to be equal to the value of the
top mass. In fact the potential is scale independent since explicit
dependence on the scale is compensated by the implicit dependence of
the parameters renormalized at this scale. $V_0$ is the value of the
potential at $H_1=H_2=0$, which has to be subtracted in order to keep
the scale invariance~\cite{Jones}.

The field dependent squarks masses are given by the eigenvalues of the
mass-squared matrix

\begin{equation}
\label{6}\left(
\begin{array}{cc}
m_Q^2+h_t^2\left| H_2^0\right| ^2+(g^2-\frac 13g^{\prime 2})(\left|
H_1^0\right| ^2-\left| H_2^0\right| ^2)/4 & h_t(A_tH_2^0+\mu H_1^{0*})\\
h_t(A_tH_2^{0*}+\mu H_1^0) & m_U^2+h_t^2\left| H_2^0\right| ^2+\frac
13g^{\prime 2}(\left| H_1^0\right| ^2-\left| H_2^0\right| ^2)
\end{array}
\right) ,
\end{equation}
where $A_t$ is the conventional trilinear soft supersymmetry breaking
parameter, $\mu $ is the Higgs mixing parameter, and the top mass
squared is given by $h_t^2\left| H_2^0\right| ^2$. The so-called
D-terms give contribution proportional to the gauge couplings and will
be ignored hereafter in order to gain approximate scale independence of
the potential, since we are including only top-stop contributions to
$\Delta V$.

Then the eigenvalues of the matrix (\ref{6}) are

\begin{equation}
\label{7}\tilde m_{t_{1,2}}^2=m_t^2+\frac 12\left[ m_Q^2+
m_U^2\pm \sqrt{(m_Q^2-m_U^2)^2+4m_t^2(A_t+\mu \cot \beta )^2}\right] ,
\end{equation}
where $\tan \beta =\left| H_2^0\right| /\left| H_1^0\right| $.

The scale independence of the effective potential (\ref{5}) is given by
the renormalization group equation

\begin{equation}
\label{8}Q^2\frac d{dQ^2}V=\left(Q^2\frac \partial {\partial Q^2}+\beta
_{m_i}\frac \partial {\partial m_i^2}+\gamma _1H_1\frac \partial
{\partial H_1}+\gamma _2H_2\frac \partial {\partial H_2}\right) V=0,
\end{equation}
where $m_i$ are the parameters of the tree-level potential and $\beta
$'s denote their $\beta $ functions. According to the approximation
mentioned above, we have ignored the scale dependence of the gauge
couplings.

The general solution of eq.(7) has the form:

$$\hat V = V(m_i(t),H_i(t),t=0)$$
where the function $V(m,H,t)$ is the perturbative expression
with the scale chosen arbitrary, $m_i$ -- are the effective
parameters of the potential,

$$H_i(t)=H_i\xi_i^{1/2}(t), \ \ \xi _i= \exp{\int\limits_{0}^{t}\gamma
_i(t')dt'}, \ \ \ t=ln\frac{M_{X}^{2}}{Q^2}$$
and $\gamma$'s are anomalous dimensions of the fields $H_i$.
In particular one can choose $Q^2$ to be equal to the top mass, as is
usually done.  However, the main distinction is whether we choose the
top mass to be field dependent and put it equal to its numerical value
after the minimization of the potential, or we take its numerical value
at the very beginning and then minimize the potential.  In the latter
case we incorporate the perturbative corrections to the potential,
while in the first case we sum all the leading logs via the
renormalization group equation.  Proceeding the first way we get

\begin{eqnarray}
V(H_1,H_2) &=& m^2_1(m_t)|H_1|^2+m^2_2(m_t)|H_2|^2
-m^2_3(m_t)(H_1H_2+h.c.) \nonumber \\
& + & \frac{g^2+g^{\prime 2}}{8}(|H_1|^2-|H_2|^2)^2 \label{9} \\
&+ &\frac{3}{32\pi^2}\left[ \tilde{m}_{t1}^4 ( \ln
\frac{\tilde{m}_{t1}^2}{m_t^2}-\frac{3}{2}) +\tilde{m}_{t2}^4 ( \ln
\frac{\tilde{m}_{t2}^2}{m_t^2}-\frac{3}{2}) + 3{m}_{t}^4 \right.
\nonumber\\
& - & \left. \tilde{m}_{Q}^4 ( \ln\frac{\tilde{m}_{Q}^2}{m_t^2}-
\frac{3}{2}) +\tilde{m}_{D}^4 ( \ln \frac{\tilde{m}_{Q}^2}{m_t^2}-
\frac{3}{2})\right] \nonumber,
\end{eqnarray}
where $m_t$ is the field dependent mass and $m_Q$ and $m_D$ are the stop
masses boundary values when $H_1=H_2=0$.

Eq.\ref{9} is the RG improved expression for the one-loop effective
potential which corresponds to taking into account all the leading log
contributions proportional to the top Yukawa coupling from all the
loops.

\newpage

\section{ Minimization of the Potential}

To find the vacuum expectation values of the Higgs fields we have to
minimize the potential. Using the notation

$$
\langle H_1^0 \xi _{1}^{1/2}\rangle =v_1,\ \langle H_2^0 \xi
_{2}^{1/2}\rangle =v_2,\ v^2=v_1^2+v_2^2,\ \tan \beta =v_2/v_1
$$
and keeping only the terms of the first order in coupling constants the
minimum of the potential (\ref{9}) is given by

\begin{eqnarray}
v^2&=&\frac{\displaystyle 4 }{\displaystyle (g^2+g^{\prime 2})
(\tan^2\beta -1)(1+\varepsilon_1)}\Bigg\{ m_1^2(1+\gamma_1)
-m_2^2(1-\gamma_1)\tan^2\beta +\varepsilon_2 \tan^2\beta \nonumber \\
&-&  \frac{3h_t^2}{16\pi^2}\left[[f(\tilde{m}^2_{t1})+
f(\tilde{m}^2_{t2})+2m^2_t]\tan^2\beta+(A_t^2\tan^2\beta -\mu^2)
\frac{f(\tilde{m}^2_{t1})-f(\tilde{m}^2_{t2})}{\tilde{m}^2_{t1}-
\tilde{m}^2_{t2}}\right]\Bigg\}, \label{10} \\
2m_3^2&=&\frac{\sin 2\beta}{1+\varepsilon_1}\Bigg\{m_1^2(1-\gamma_2)
+m_2^2(1-\gamma_1)-\varepsilon_2 \label{11} \\
&+&  \frac{3h_t^2}{16\pi^2}\left[f(\tilde{m}^2_{t1})
+f(\tilde{m}^2_{t2})-2f(m^2_t)+(A_t  +\mu\tan\beta)(A_t +\mu\cot\beta)
\frac{f(\tilde{m}^2_{t1})-f(\tilde{m}^2_{t2})}{\tilde{m}^2_{t1}-
\tilde{m}^2_{t2}}\right]\Bigg\} ,\nonumber
\end{eqnarray}
where

\begin{eqnarray*}
f(m^2)&=&m^2(\log\frac{m^2}{m_t^2}-1) , \\
\varepsilon_1&=&\gamma_1\cos^2\beta-\gamma_2\sin^2\beta , \\
\varepsilon_2&=&3\left[(\tilde{\alpha}_2M_2^2
+\frac{1}{5}\tilde{\alpha}_1M_1^2+\tilde{\alpha}_2\mu^2+
\frac{1}{5}\tilde{\alpha}_1\mu^2)(\cot^2\beta+1)+2(\tilde{\alpha}_2M_2+
\frac{1}{5}\tilde{\alpha}_1M_1)\mu\cot\beta\right] , \\
\gamma_1&=&-\frac{3}{2}(\tilde{\alpha}_2+
\frac{1}{5}\tilde{\alpha}_1) , \\
\gamma_2&=&\frac{3}{2}(2Y_t-\tilde{\alpha}_2
-\frac{1}{5}\tilde{\alpha}_1) , \\
Y_t&=&\frac{h_t^2}{16\pi^2} ,
\tilde{\alpha}_i=\frac{\alpha_i}{4\pi}=\frac{g_i^2}{16\pi^2}.
\end{eqnarray*}

Here the values of all the mass parameters and couplings are taken at
the scale equal to the top mass.

\newpage

\section{Corrections to the Higgs Masses}

Having in mind eqs.(\ref{11}) we are now in a position to calculate the
RG improved radiative corrections to the masses.

One has:

\begin{equation}
\label{12}M_Z^2=2\frac{\displaystyle m_1^2(1+\gamma _1)-m_2^2(1-\gamma
_1)\tan ^2\beta -\tilde \Delta _Z^2}{(\tan {}^2\beta
-1)(1+\varepsilon _1)},
\end{equation}
where
$$\tilde \Delta _Z^2=\Delta _Z^2-\varepsilon _2\tan {}^2\beta $$
and $\Delta _Z^2$ is the one-loop radiative correction~\cite{Zwirner},
\cite{de Boer}

\begin{equation}
\label{13}\Delta _Z^2=\frac{3g^2}{32\pi ^2}\frac{m_t^2}{M_W^2\cos
{}^2\beta }\left[ f(\tilde m_{t1}^2)+f(\tilde
m_{t2}^2)+2m_t^2+(A_t^2m_0^2-\mu ^2\cot {}^2\beta )\frac{f(\tilde
m_{t1}^2)-f(\tilde m_{t2}^2)}{\tilde m_{t1}^2-\tilde m_{t2}^2}\right] .
\end{equation}

Using eqs.(\ref{12}) we can also calculate the corrections to the
squared masses in the CP-odd neutral sector. Just like in the usual
case~\cite {Zwirner} taking the second derivative of the full potential
with respect to $\phi _i\equiv ImH_i^0$ one has:
\begin{equation}
\label{14}\left.\left(\frac{\partial^2V}{\partial\phi_i\partial\phi_j}
\right) \right| _{v_1,v_2}=\left(
\begin{array}{cc}
\tan \beta & 1 \\
1 & \cot \beta
\end{array}
\right) \Delta ,
\end{equation}
where
$$\Delta=2m_3^2-\frac{3g^2}{32\pi ^2\sin {}^2\beta}\frac{m_t^2}{M_W^2}
Am_0\mu \frac{f(\tilde m_{t1}^2)-f(\tilde
m_{t2}^2)}{\tilde m_{t1}^2-\tilde m_{t2}^2} $$

This gives for the CP-odd Higgs mass

\begin{equation}
\label{15}m_A^2=\frac 1{1+\varepsilon_1}\left\{ m_1^2(1-\gamma
_2)+m_2^2(1-\gamma _1)+\tilde \Delta _A^2\right\} ,
\end{equation}
where
$$\tilde \Delta _A^2=\Delta _A^2-\varepsilon _2,$$
and
\begin{equation}
\label{16}\Delta_A^2=\frac{3g^2}{32\pi^2}\frac{m_t^2}{M_W^2\sin^2\beta}
\left[ f(\tilde m_{t1}^2)+f(\tilde m_{t2}^2)+2m_t^2+(A_t^2m_0^2+\mu ^2)
\frac{ f(\tilde m_{t1}^2)-f(\tilde m_{t2}^2)}{\tilde m_{t1}^2-
\tilde m_{t2}^2} \right]
\end{equation}

For the CP-even sector we have to differentiate the potential with
respect to $\psi _i=ReH_i^0.$ We find

\begin{equation}\label{a}
\left.\left(\frac{\partial^2V}{\partial\psi_i\partial\psi_j}\right)
\right| _{v_1,v_2}=\left(
\begin{array}{cc}
\cot \beta  & -1-\delta _1 \\
-1-\delta _1 & \tan \beta (1+\delta _2)
\end{array}
\right) M_Z^2\sin 2\beta  \end{equation} $$ +\left(
\begin{array}{cc}
\tan \beta  & -1-\delta _3 \\
-1-\delta _3 & \cot \beta (1+\delta _4)
\end{array}
\right) \Delta   +2\left(
\begin{array}{cc}
\Delta _{11} & \Delta _{12} \\
\Delta _{12} & \Delta _{22}
\end{array}
\right) +2\left(
\begin{array}{cc}
0 & -\delta _5 \\
-\delta _5 & \delta _6
\end{array}
\right) , $$
where

\begin{eqnarray*}
\delta_1&=&\gamma_1\cot^2\beta-\gamma_2, \\
\delta_2&=&(\gamma_1+\gamma_2)(\cot^2\beta-1), \\
\delta_3&=&2\gamma_1, \\
\delta_4&=&2\gamma_2, \\
\delta_5&=&6
\left[(\tilde{\alpha}_2M_2^2+\frac{1}{5}\tilde{\alpha}_1M_1^2+
\tilde{\alpha}_2\mu^2+\frac{1}{5}\tilde{\alpha}_1\mu^2)\cot^2\beta
+2(\tilde{\alpha}_2M_2+\frac{1}{5}\tilde{\alpha}_1M_1)\mu\right] , \\
\delta_6&=&6
\left[(\tilde{\alpha}_2M_2^2+\frac{1}{5}\tilde{\alpha}_1M_1^2+
\tilde{\alpha}_2\mu^2+\frac{1}{5}\tilde{\alpha}_1\mu^2)(\cot^2\beta-1)
+2(\tilde{\alpha}_2M_2+\frac{1}{5}\tilde{\alpha}_1M_1)
\mu\cot\beta \right. \\
 && \left. -Am_0\mu Y \cot\beta \right]
\end{eqnarray*}
and $\Delta_{ij}$'s are ~\cite{Zwirner}:
\begin{eqnarray*}
\Delta_{11}&=&\frac{3g^2}{16\pi^2}\frac{m^4_t}{\sin^2\beta M^2_W}
\left[\frac{\mu(A_t m_0+
\mu\cot\beta )}{\tilde{m}^2_{t1}-\tilde{m}^2_{t2}}
\right]^2d(\tilde{m}^2_{t1},\tilde{m}^2_{t2}), \\
\Delta_{22}&=&\frac{3g^2}{16\pi^2}\frac{m^4_t}{\sin^2\beta M^2_W}\left[
\ln (\frac{\tilde{m}^2_{t1}\tilde{m}^2_{t2}}{m^4_t})+
\frac{2A_t m_0 (A_t m_0+
\mu\cot\beta )}{\tilde{m}^2_{t1}-\tilde{m}^2_{t2}} \ln
(\frac{\tilde{m}^2_{t1}}{\tilde{m}^2_{t2}}) \right.\\
 & & \left.+\left[
 \frac{A_t m_0(A_t m_0 +\mu\cot\beta
)}{\tilde{m}^2_{t1}-\tilde{m}^2_{t2}}\right]^2
d(\tilde{m}^2_{t1},\tilde{m}^2_{t2})\right], \\
\Delta_{12}&=&\frac{3g^2}{16\pi^2}\frac{m^4_t}{\sin^2\beta M^2_W}
\frac{\mu(A_t m_0 +
\mu\cot\beta )}{\tilde{m}^2_{t1}-\tilde{m}^2_{t2}} \left[
\ln (\frac{\tilde{m}^2_{t1}}{\tilde{m}^2_{t2}}) +
\frac{A_t m_0 (A_t m_0+\mu\cot\beta
)}{\tilde{m}^2_{t1}-\tilde{m}^2_{t2}}
d(\tilde{m}^2_{t1},\tilde{m}^2_{t2})\right] \\
\end{eqnarray*}
$$ h(m^2)=\frac{m^2}{m^2-\tilde{m}^2_{q}}\ln \frac{m^2}{\tilde{m}^2_{q}}
\;\;\;\;\; d(m^2_1,m^2_2)=2-\frac{m^2_1+m^2_2}{m^2_1-m^2_2}\ln
\frac{m^2_1}{m^2_2},$$
$\tilde{m}^2_q$ is the mass of a light squark.

The diagonalization of the matrix (\ref{a}) gives us the masses of
the CP-even neutral Higgses.

For the charged ones one has the usual expression~\cite{Zwirner}

\begin{equation}
m_{H^{\pm}}^2=m_A^2+M_W^2+\Delta _H^2,
\end{equation}
where

\begin{equation}
\Delta _H^2=-\frac{3g^2}{32\pi^2}\frac{m_t^4\mu^2}{\sin {}^4\beta M_W^2}
\frac{h(\tilde m_{t1}^2)-h(\tilde m_{t2}^2)}{\tilde m_{t1}^2
-\tilde m_{t2}^2}
\end{equation}
and the only difference is that $m_A$ is given by eq.(\ref{15}).

\section{Results and Conclusion}

Resulting expressions for the Higgs masses differ from those obtained
without RG summation. To calculate the corrections one has to perform
the usual procedure~\cite{Ross},~\cite{Ibanez},~\cite{de Boer} of
fitting the set of soft breaking parameters, $m_0,m_{1/2},\mu ,\tan
\beta ,A,$ as well as the top mass $m_t$. This is not a straightforward
operation, since one has to fulfill many requirements simultaneously,
thus defining the optimized best fit~\cite{de Boer}. Having performed
the complete procedure with the help of computer programm we have got
the best fit values of the parameters mentioned above and used them in
our formufae.

One of the important observation is the character of $m_t$ dependence
of the results. Comparing with that obtained without account of the
RG summation~\cite{de Boer}, we find it to be smoother. Fig.1 shows
our results.

One should note, that $m_t$ in our formulae is the running top mass,
which is connected with the physical (pole) top mass by the
relation~\cite{100}:

$$ M_t^{pole}=m_t\left(1+\frac{4}{3}\frac{\alpha _s}{\pi }\right)
\approx 1.06 m_t $$

We conclude that the account of RG summation procedure can introduce the
changes in the predictions of the Higgs masses. One can observe from
Fig.1 that there exist the lower bound on the Higgs mass of about 95
GeV. In the interval of top mass preferable according to the recent
CDF~\cite{CDF} data ($M_t^{pole}=174\pm 10\; GeV$), that corresponds to
the running top mass of 164 $GeV$, the $m_t$ dependence of the
lightest Higgs mass is very weak and $h$ appears to be lighter than 100
GeV.

\vglue 2cm

\subsection*{Acknowledgements}
The numerical analysis was performed with the help of the computer
program, developed in~\cite{de Boer}. We are grateful to W.de Boer and
R.Ehret from Karlshrue University for valuable discussions and for
necessary modification of the program.

\newpage

\begin{figure}[ht]
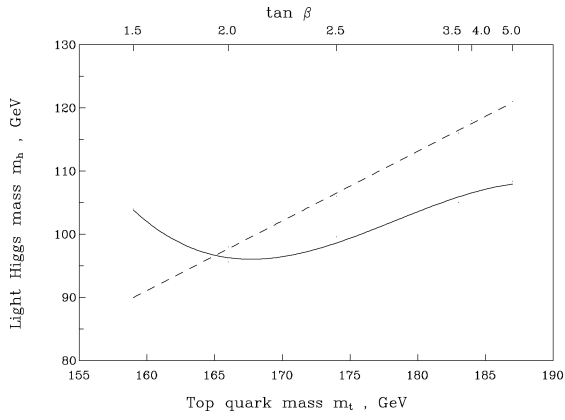

\vspace{.0cm}
\special{em:graph effpot.msp}
\vspace{12cm}
\caption{$m_t$ dependence of the light Higgs boson mass for
$m_0=400\; GeV$, $m_{1/2}=200\; GeV$. Dashed line corresponds to the
1-loop approximation, solid one corresponds to the RG improved
radiative corrections.}
\label{fig1}
\end{figure}

\end{document}